\documentclass[preprint,12pt]{elsarticle}

\usepackage{amssymb}
\usepackage{lineno}

\usepackage[T1]{fontenc}

\usepackage[linesnumbered,ruled,vlined]{algorithm2e}
\SetKwComment{Comment}{// }{}
\SetKwInput{KwCompute}{compute}

\usepackage{float}
\newfloat{lstfloat}{t}{lop}

\usepackage{hyperref}

\usepackage{xcolor}
\definecolor{mygrey}{rgb}{0.33,0.33,0.33}

\usepackage[lighttt]{lmodern}

\usepackage{listings}

\usepackage{rotating}

\usepackage{subfig}

\def\imagecentered#1{$\vcenter{\null\hbox{#1}}$}

\newcommand{\martinezFigLabel}[6]{
  \begin{tabular}{@{}cc@{}}
    \begin{minipage}{0cm}
      \begin{sideways} \scriptsize #5~#6 \end{sideways}
    \end{minipage}
    &
    \imagecentered{\includegraphics[width=#1]{#2}}
    \tabularnewline
    &
    \scriptsize #3~#4
    \tabularnewline
  \end{tabular}
}

\newcommand{\martinezFigLabelTwo}[4]{
  \begin{tabular}{@{}cc@{}}
    \begin{minipage}{0cm}
      \begin{sideways} \scriptsize #3~#4 \end{sideways}
    \end{minipage}
    &
    \imagecentered{\includegraphics[width=#1]{#2}}
  \end{tabular}
}

\newcommand{\figSrc}[1]{./figures/#1}
\newcommand{\codeSrc}[1]{./codes/#1}

\newcommand{\checkPM}[1]{#1}                % Dummy check
\newcommand{\checkPMB}[1]{#1}               % Dummy check

\journal{Journal of Parallel and Distributed Computing}

\begin{document}

\begin{frontmatter}



\title{Improving the performance of classical linear algebra iterative methods via hybrid parallelism\tnoteref{copyright,doi}}
\newpageafter{title} % Required by Journal of Parallel and Distributed Computing

\tnotetext[copyright]{\textcopyright 2023. This manuscript version is made available under the CC BY-NC-ND 4.0 license \color{blue}{\url{https://creativecommons.org/licenses/by-nc-nd/4.0/}}}
\tnotetext[doi]{Published journal article available at \color{blue}{\url{https://doi.org/10.1016/j.jpdc.2023.04.012}}}


\author[label1,label2]{Pedro J.~Martinez-Ferrer\corref{cor1}}
\ead{pedro.martinez.ferrer@upc.edu}

\affiliation[label1]{organization={Departament d'Arquitectura de Computadors (DAC), Universitat Politècnica de Catalunya - BarcelonaTech (UPC)},%Department and Organization
            addressline={Campus Nord, Edif. D6, C. Jordi Girona 1-3},
            city={08034 Barcelona},
            country={Spain}}

\affiliation[label2]{organization={Barcelona Supercomputing Center (BSC)},%Department and Organization
            addressline={Pl. Eusebi Güell 1-3},
            city={08034 Barcelona},
            country={Spain}}

\author[label2]{Tufan Arslan}
\ead{tufan.arslan@bsc.es}

\author[label2]{Vicenç Beltran}
\ead{vbeltran@bsc.es}

\cortext[cor1]{Corresponding author}

\begin{abstract}
We propose fork-join and task-based hybrid implementations of four
classical linear algebra iterative methods (Jacobi, Gauss--Seidel,
conjugate gradient and biconjugate gradient \checkPM{stabilized})
\checkPM{on CPUs} as well as variations of them.  \checkPM{This class
  of algorithms, that are ubiquitous in computational frameworks,} are
duly documented and the corresponding source code is made publicly
available for reproducibility.  Both weak and strong scalability
benchmarks are conducted to statistically analyse their relative
efficiencies.

The weak scalability results assert the superiority of a task-based
hybrid parallelisation over MPI-only and fork-join hybrid
implementations.  Indeed, the task-based model is able to achieve
speedups of up to 25\% larger than its MPI-only counterpart depending
on the numerical method and the computational resources used.  For
strong scalability scenarios, hybrid methods based on tasks remain
more efficient with moderate computational resources where data
locality does not play an important role.  Fork-join hybridisation
often yields mixed results and hence does not \checkPM{seem to bring}
a competitive advantage over a much simpler MPI approach.

\end{abstract}
\newpageafter{abstract} % Required by Journal of Parallel and Distributed Computing

%% \begin{graphicalabstract}
%% %\includegraphics{grabs}
%% \end{graphicalabstract}

%% \begin{highlights}
%% \item Four classical linear algebra iterative methods are hybridised \checkPM{on CPUs}.
%% \item Implementations with MPI, fork-join, and task-based parallel models are compared.
%% \item For weak scalability scenarios, tasks yield the best performance results.
%% \item For strong scalability scenarios, tasks remain competitive with moderate resources.
%% \item Fork-join hybrid methods \checkPM{often yield mixed results}.
%% \end{highlights}


\end{frontmatter}

%% \modulolinenumbers[5]


\section{Introduction}
\label{sec:introduction}
Numerical linear algebra plays an important role in the resolution of
many scientific and engineering problems.  It emerges naturally from
the application of the finite difference, finite volume and finite
element methods to the numerical resolution of partial differential
equations (PDEs) often yielding sparse linear systems.  There exist
direct and iterative linear algebra methods, being this second class
of techniques computationally preferable for large and sparse
systems~\citep{Freund1992}.  The first iterative methods were based on
relaxation of the coordinates and the ones that remain quite popular
even today are the Jacobi and Gauss--Seidel (GS) algorithms.  The GS
method constitutes an improvement over Jacobi due to its ability to
correct the current approximate solution.  It is worth mentioning
that, in practice, these two techniques are seldom used separately and
hence they are often combined (via preconditioning) with more
efficient and sophisticated Krylov subspace methods (KSMs).  Two great
KSM exponents are the conjugate gradient (CG) and the biconjugate
gradient \checkPM{stabilized} (BiCGStab) methods.  The former can be
regarded as one of the best-known iterative algorithms for solving
sparse symmetric linear systems whilst the latter is made extensible
to \checkPM{nonsymmetric} linear systems~\citep{Saad2003}.  It is
therefore not uncommon to find these four algebraic methods integrated
in many numerical libraries and frameworks~\citep{Weller1998}.

The \checkPM{three types} of arithmetic operations involved in the
aforementioned iterative methods: \checkPM{(i)} matrix--vector
multiplications, \checkPM{(ii)} vector updates, \checkPM{and (iii)}
scalar products, are well suited for both vector and parallel
computers~\citep{Dongarra1991}.  Matrices and vectors can be
efficiently distributed among the processors but the global
communication caused, for instance, by the inner product of two
vectors represents a major bottleneck on distributed computers.  The
performance penalty associated with collective communications
aggravates with the number of parallel processes and hence can
dramatically affect the \checkPM{scalability} of iterative methods.
This has motivated many researchers to find solutions aimed at
reducing the impact of synchronisation barriers present in iterative
methods.

In addition to this, parallel implementations of linear algebra
methods \checkPM{tend to rely primarily} on the message passing
interface (MPI) library~\citep{Walker1994} thus leaving other parallel
paradigms and approaches off the
table~\citep{Weller1998,Dongarra1991}.  We are referring to the
popular OpenMP fork-join parallelism~\citep{Dagum1998} as well as the
less-known task dependency archetype brought by the OmpSs programming
model~\citep{Perez2017} that was \checkPM{first} integrated into
OpenMP \checkPM{3.0 in 2008}.  \checkPM{Having mentioned all that,
  there exist computational frameworks with support for hybrid
  parallel programming \citep{Guo2015} and also numerical libraries
  such as PETSc~\citep{petsc1997},
  Trilinos~\citep{trilinos2020}\checkPMB{,} or Hypre~\citep{hypre2011}
  proposing several levels of parallelism, both hybrid (typically MPI
  with OpenMP fork-join) and heterogeneous (GPUs).  What is more,}
although \checkPM{task-based} hybrid implementations of iterative
algebraic methods do exist, e.g.\ \citet{Aliaga2016} or
\citet{Zhuang2017}, they still remain scarce in the scientific
literature and, to the best of our knowledge, are not present in any
major computational frameworks.  The present study contributes to
enhancing the parallel performance of the four aforementioned
numerical methods \checkPM{on CPUs by}:
\begin{itemize}
\item Proposing algorithm variations aimed at reducing or eliminating
  communication barriers \checkPM{via OpenMP and OmpSs-2 tasks}.
\item Bringing hybrid parallelism via fork-join and task-based
  shared-memory programming models in conjunction with MPI.
\item Carrying out systematic benchmarks and statistical analysis of
  the hybrid implementations and comparing them against MPI-only
  parallel code.
\end{itemize}

The remainder of this paper is organised as follows.
Section~\ref{sec:related-work} presents the related work and
Section~\ref{sec:numerical-methods} describes the numerical methods as
well as their hybrid implementations.
Section~\ref{sec:results-and-discussion} contains the main findings of
this work and discusses the results in detail.  Finally,
Section~\ref{sec:conclusions-future-work} gives the main conclusions
and proposes some future work.

\section{Related work}
\label{sec:related-work}
Countless examples attempting to improve the parallel efficiency of
iterative linear algebra methods can be found in the literature.
Skipping over the straightforward Jacobi algorithm, both bicolouring
(e.g.\ a red--black colour scheme)~\citep{Fox1989} and
multicolouring~\citep{Golub1993} techniques have been proposed for
providing independence and therefore some degree of parallelism to the
Gauss--Seidel algorithm.  However, the appropriateness of such
colouring schemes largely depends on the coefficient matrix associated
with the linear system.  What is more, in order to prevent
parallelisation difficulties, \emph{processor-} and
\emph{thread-localised} GS methods are often employed instead of a
\emph{true} GS parallel method able to maintain the exact same
convergence rate of the sequential algorithm~\cite{Shang2009}.

As for Krylov subspace methods, one of the most common improvements
consists of reducing the number of global reductions associated with
the calculation of the inner product of two vectors.  An earlier
example applied to the CG algorithm can be found in the book of
\citet{Barrett1994}.  Other authors have successfully reduced the
total number of global scalar products to just one via $s$-step
methods~\citep{Chronopoulos1989}.  \checkPM{One of the main drawbacks
  of} this approach is that numerical issues start arising for
increasing orthogonal directions represented by $s$.  Other
alternatives seek to overlap computational kernels with communication:
the pipelined versions suggested by~\citet{Ghysels2014} result in an
asynchronous preconditioned CG algorithm where one of global
reductions is overlapped with the sparse matrix--vector multiplication
(SpMV) kernel and the other takes advantage of the preconditioning
step to hide its communication latency.  A more contemporary approach
with the name of ``iteration-fusing'' attempts to partially overlap
iterations by postponing the update of the solution residual at the
end of each iteration~\cite{Zhuang2017}.  The main disadvantage of
this method is to determine how often to check for the residual as it
requires an exhaustive search and remains problem-specific.  Needless
to say, all the aforementioned works only exploit parallelism via MPI
with the exception of the iteration-fusing variant of
\citet{Zhuang2017}, which heavily relies upon OpenMP tasking to
achieve overlapping.  This method has been recently
improved \cite{Barreda2019} by intertwining computation and
communication tasks using the OmpSs-2 task-based programming model and
the task-aware MPI (TAMPI) library~\citep{Sala2019}.

The BiCGStab method has also received a lot of attention.  For
instance, the MBiCGStab variant proposed by \citet{Jacques1999}
reduced the three original global synchronisations to only two.  Based
on the same concept, the nonpreconditioned BiCGStab was improved by
\citet{Yang2002} with a single global communication that could be
effectively overlapped with computation.  Other authors have proposed
reordered and pipelined BiCGStab methods, all of them taking advantage
of the preconditioning step to hide the communication latency.  It is
worth noting that, although these variants do not change the numerical
stability of the method, they do incur an extra number of computations
(due to the increasing role of vector operations) with respect to the
classical algorithm.  Consequently, there is a trade-off between
communication latency and extra computation and therefore it becomes
tricky to determine a priori the suitability of each method over the
classical BiCGStab algorithm, as demonstrated by
\citet{Krasnopolsky2020}.  What is more, such a comparison will depend
on whether parallelisation relies exclusively on the MPI
distributed-memory model as in~\citep{Krasnopolsky2020} or it is
combined with shared-memory parallelism.

To the best of our knowledge, there are not bibliographic references
with an extensive comparison of several algebraic methods following
pure and hybrid MPI paralleli\checkPMB{s}ation strategies
\checkPM{using OpenMP and/or OmpSs-2 tasks}.  To this end, the present
study sheds light on the potential gains that hybrid applications can
achieve.  At the same time, this work is also intended as a reference
for effective hybrid programming and code implementation that future
researchers can adopt when developing highly scalable numerical
methods for linear algebra.

\section{Numerical methods and parallel implementations}
\label{sec:numerical-methods}
This section describes all the numerical methods presented in this
work as well as their hybrid, parallel implementations currently based
on a combination of MPI with OpenMP and OmpSs-2.

\subsection{Description of the algorithms}
\label{sec:description-algorithms}
We are particularly interested in bringing hybrid parallelism to
linear algebra iterative methods for the numerical resolution of
systems of equations of the form ${\bf A} \cdot {\bf x} = {\bf b}$,
where ${\bf A}$ represents a known sparse matrix, ${\bf b}$ is a known
array, and ${\bf x}$ refers to an unknown vector satisfying the
equality.  Such sparse linear systems typically \checkPMB{arise} from
the numerical approximation of PDEs related to engineering problems
like those found in computational fluid dynamics (CFD).  We place
special emphasis on the OpenFOAM library~\citep{Weller1998}.  This
open-source CFD framework provides an extensive range of features to
solve complex fluid flows and, because it remains free to use, has a
large user base across most areas of engineering and science, from
both commercial and academic organisations.  \checkPM{We have
  identified within this CFD library the four most commonly used
  iterative methods worth studying.}

There is a plethora of numerical methods that can be applied to the
kind of sparse systems described in the previous paragraph.  Some of
them can even be combined with preconditioning or smoothing
techniques, which are not covered in this work.  The two most basic
methods that we address are the Jacobi method and the \emph{symmetric}
Gauss--Seidel method.  The latter applies one forward sweep followed
by a backward sweep within the same iteration and can be seen as an
improvement of the former since the computation of a given element of
the array ${\bf x}$ uses all the elements that have been previously
computed in the current iteration.  The other two numerical algorithms
covered in this work are CG and BiCGStab.  These KVMs rely primarily
on SpMV calculations and can achieve better convergence rates than
Jacobi and Gauss--Seidel~\citep{Saad2003}.  Since these four
algorithms remain highly popular, and are extensively used in
OpenFOAM, they become great exponents to study different hybrid
parallelisation strategies.

\begin{algorithm}[tbh]
\caption{Conjugate gradient, nonblocking algorithm (CG-NB) \checkPM{with a schematic task implementation (\texttt{Tk})}.}\label{alg:CG-NB}
{${\bf r}_0 = {\bf b} - {\bf A} \cdot {\bf x}_0$, ${\bf p}_0 = {\bf r}_0$, $\alpha_0 = \alpha_{{\rm n},0} / \alpha_{{\rm d},0} = {\bf r}_0 \cdot {\bf r}_0 / ( ({\bf A} \cdot {\bf p}_0) \cdot {\bf p}_0 )$} \;
\For{$j = 1, 2, \ldots$}{
  {\bf if} $\sqrt{\alpha_{{\rm n},j-1}} < \epsilon$ {\bf then} break \Comment*[l]{Exit loop}
  ${\bf r}_j = {\bf r}_{j-1} - \alpha_{j-1} {\bf A} \cdot {\bf p}_{j-1}$ \Comment*[r]{Tk 0}
  $\alpha_{{\rm n},j} = {\bf r}_j \cdot {\bf r}_j$ \Comment*[r]{Tk 0}
  ${\bf A} \cdot {\bf p}_j = {\bf A} \cdot {\bf r}_j + \frac{\alpha_{{\rm n},j}}{\alpha_{{\rm n},j-1}} {\bf A} \cdot {\bf p}_{j-1}$ \Comment*[r]{Tk 1 \& 2}
  ${\bf p}_j = {\bf r}_j +  \frac{\alpha_{{\rm n},j}}{\alpha_{{\rm n},j-1}} {\bf p}_{j-1}$ \Comment*[r]{Tk 2}
  $\alpha_{{\rm d},j} = ({\bf A} \cdot {\bf p}_j) \cdot {\bf p}_j$ \Comment*[r]{Tk 2}
  ${\bf x}_j = {\bf x}_{j-1} - \frac{\alpha^2_{{\rm n},j-1}}{\alpha_{{\rm d},j-1} \cdot \alpha_{{\rm n},j}} ({\bf p}_{j} - {\bf r}_{j})$  \Comment*[r]{Tk 3}
}
{${\bf x} = {\bf x}_{\rm l} - \alpha_{\rm l} {\bf p}_{\rm l}$} \;
\end{algorithm}

In this work, we propose for hybridisation two
\emph{nonpreconditioned} KVMs, namely CG-NB and BiCGStab-B1, described
by Algorithms~\ref{alg:CG-NB} and \ref{alg:BiCGStab-B1}, respectively.
On the one hand, the CG-NB method does apply the SpMV kernel on vector
${\bf r}_j$ so that the product ${\bf A} \cdot {\bf p}_j$ is simply
computed as an array update of ${\bf A} \cdot {\bf r}_j$ (line 6 of
Algorithm~\ref{alg:CG-NB}).  This removes the two communication
barriers of the original CG algorithm thus rendering it truly
\emph{nonblocking}: note that the collective reduction caused by the
dot product (line 5) can be overlapped with the SpMV on ${\bf r}_j$ to
then proceed with line 6.  Although this algorithm is arithmetically
equivalent to the classical one, it might converge slightly different.

\begin{algorithm}[tbh]
\caption{Biconjugate gradient \checkPM{stabilized}, one blocking algorithm (BiCGStab-B1) \checkPM{with a schematic task implementation (\texttt{Tk})}.}\label{alg:BiCGStab-B1}
{${\bf r}_0 = {\bf b} - {\bf A} \cdot {\bf x}_0$, ${\bf p}_0 = {\bf r}_0$, $\checkPM{\beta_0} = \mathbf{r}_0 \cdot \mathbf{r}_0$, $\mathbf{r}' = \mathbf{r}_0/\sqrt{\beta_0}$, $\alpha_{{\rm n},0} = {\bf r}_0 \cdot {\bf r}'$} \;
\For{$j = 0, 1, \ldots$}{
  $\alpha_{{\rm d},j} = ({\bf A} \cdot {\bf p}_j) \cdot {\bf r}'$ \Comment*[r]{Tk 0 (blocking)}
  ${\bf s}_j = {\bf r}_j - \alpha_j {\bf A} \cdot {\bf p}_j$ \Comment*[r]{Tk 1}
  $\omega_j = ({\bf A} \cdot {\bf s}_j) \cdot {\bf s}_j / ( ({\bf A} \cdot {\bf s}_j) \cdot ({\bf A} \cdot {\bf s}_j) )$ \Comment*[r]{Tk 2}
  ${\bf x}_{j+1/2} = {\bf x}_j + \alpha_j {\bf p}_j$ \Comment*[r]{Tk 3}
  {\bf if} $\sqrt{\beta_j} < \epsilon$ {\bf then} break \Comment*[l]{Exit loop}
  ${\bf x}_{j+1} = {\bf x}_{j+1/2} + \omega_j  {\bf s}_j$ \Comment*[r]{Tk 4}
  ${\bf r}_{j+1} = {\bf s}_j - \omega_j {\bf A} \cdot {\bf s}_j$ \Comment*[r]{Tk 4}
  $\alpha_{{\rm n},j+1} = {\bf r}_{j+1} \cdot {\bf r}'$ \Comment*[r]{Tk 4}
  $\beta_{j+1} = {\bf r}_{j+1} \cdot {\bf r}_{j+1}$ \Comment*[r]{Tk 4}
  ${\bf p}_{j+1/2} = {\bf p}_j - \omega_j {\bf A} \cdot {\bf p}_j$ \Comment*[r]{Tk 5}
  \uIf(\Comment*[h]{Restart procedure}){$\sqrt{\alpha_{{\rm n},j+1}} < \varepsilon$}{
    ${\bf p}_{j+1} = {\bf r}_{j+1}$ \Comment*[r]{Tk 6}
    ${\bf r}' = {\bf r}_{j+1}/\sqrt{\beta_{j+1}}$ \Comment*[r]{Tk 6}
  }
  \Else(\Comment*[h]{Regular procedure}){
    ${\bf p}_{j+1} = {\bf r}_{j+1} + \frac{\alpha_{{\rm n},j+1}}{\alpha_{{\rm d},j} \cdot \omega_j} {\bf p}_{j+1/2}$ \Comment*[r]{Tk 7}
  }
}
{${\bf x} = {\bf x}_{\rm l} + \omega_{\rm l} {\bf s}_{\rm l}$} \;
\end{algorithm}

On the other hand, the BiCGStab-B1 variant permutes the order of
certain operations in order to eliminate two of the three barriers
present in the original algorithm.  It is the dot product associated
with $\alpha_{{\rm d},j}$ (line 3 of Algorithm~\ref{alg:BiCGStab-B1})
the responsible of such blocking synchronisation.  However,
\checkPMB{the two scalar products, i.e.\ numerator and denominator of
  $\omega_j$ (line 5), can be overlapped with the update of ${\bf
    x}_{j+1/2}$ at line 6.  Likewise,} the scalar products
corresponding to $\alpha_{{\rm n},j+1}$ and $\beta_{j+1}$ (lines
10--11) can be overlapped with the update of ${\bf p}_{j+1/2}$ at line
12.  \checkPMB{It is worth noting that the effective overlapping of
  these two parts of the BiCGStab-B1 algorithm is only possible if the
  computation times associated with vector updates remain larger than
  those of collective communications.}

\checkPM{We highlight that the major novelty of
  Algorithms~\ref{alg:CG-NB}--\ref{alg:BiCGStab-B1} resides in their
  combination with a task-based programming language that can fully
  exploit their parallel performance.  Note that
  Algorithms~\ref{alg:CG-NB}--\ref{alg:BiCGStab-B1} showcase the
  strict minimum number of \emph{distinct} tasks necessary to fully
  parallelise them (see r.h.s.\ comments beginning with keyword
  \texttt{Tk}): four tasks for the CG-NB algorithm and eight tasks for
  BiCGStab.  In this document, the task notation has been simplified
  and enumerated for clarity and the reader is referred to
  \cite{martinez2022} for the actual task implementation in full
  detail.}

As it is the case with other strategies from the
literature~\citep{Jacques1999,Yang2002}, the two proposed methods do
incur extra operations.  In the case of CG-NB, there is an additional
vector update (line 9) which can be optimised via the ad hoc kernel
${\bf z} := a \cdot {\bf x} + b \cdot {\bf y} + c \cdot {\bf z}$ that
reuses memory.  \checkPM{Let $r$ and $\bar{n}$ represent the number of
  rows and the average number of nonzeros per row of the sparse
  matrix, respectively.}  \checkPMB{A rough estimate of the total
  number of accessed elements} per iteration of the CG-NB algorithm is
given by \checkPM{$(15+\bar{n})r$}, which is slightly larger than the
\checkPM{$(12+\bar{n})r$} corresponding to CG.  Applying the same
reasoning, one can find the exact same difference of \checkPM{$3r$}
elements between the BiCGStab algorithm, \checkPM{$(21 + 2\bar{n})r$},
and the variant proposed here, \checkPM{$(24 + 2\bar{n})r$}.
\checkPM{In this work,} we consider two sparsity levels:
\checkPM{$\bar{n}=7$} and \checkPM{$\bar{n}=27$}.  \checkPM{The lower
  value is derived from applying a 7-point centred stencil on a
  three-dimensional structured hexahedral mesh and is typical of an
  OpenFOAM application.  The higher sparsity level originates from
  applying a 27-point centred stencil to the same mesh and is actively
  used by the well-known HPCG benchmark~\citep{Dongarra2016}.}
Therefore, the \emph{maximum} relative increase of touched elements
will be given by \checkPM{$3/(12+\bar{n}) \approx 15.8\%$} for CG-NB
and \checkPM{$3/(21+2\bar{n}) \approx 8.6\%$} for BiCGStab-B1.  These
figures are on par with other approaches proposed in the past; for
example, a similar amount of extra operations is required for the
IBiCGStab algorithm proposed by \citet{Yang2002} after aggregating
every additional vector update and scalar product into a single,
\checkPM{ad hoc} function.  On the one hand, this improved algorithm
of Yang \& Brent yields a single, nonblocking synchronisation; on the
other hand, it relies upon the transpose of the input matrix and
necessitates much more system memory.  Likewise, it does not offer
support for restarting but, if it were the case, then it would require
a third SpMV operation per iteration affecting severely its
performance.  Restarting not only helps improving numerical
convergence, but it is also critical for task-based parallel
implementations (see Section~\ref{sec:parallelisation-CG-NB}).  For
the reasons mentioned above, we cannot consider the IBiCGStab
algorithm suitable for hybridisation.

\subsection{Description of the parallel framework}
\label{sec:parallel-framework}
Even though OpenFOAM already provides a MPI parallel implementation of
the classical versions of the methods described in the previous
section, it is not used here as a reference framework due to its
tremendously large code base.  Instead, we implement our numerical
codes within the HPCCG benchmark~\cite{Heroux2017}.  This application
is a minimalist version of the famous HPCG
benchmark~\cite{Dongarra2016} used in the TOP500 supercomputer
list\footnote{https://www.top500.org} and was developed by the same
authors in the context of the Mantevo project~\cite{Heroux2009}.
HPCCG applies the classical CG on a sparse system encoded in the
popular compressed sparse row (CSR) matrix format.  Similarly to
OpenFOAM, HPCCG only supports distributed parallelism via MPI
(MPI-only).

We provide two approaches to hybrid parallelism, one using the
\emph{fork-join} paradigm with the clause \texttt{parallel for} from
OpenMP (MPI-OMP$_{\rm fj}$) and the other based on tasks via either
OpenMP (MPI-OMP$_{\rm t}$) or OmpSs-2 (MPI-OSS$_{\rm t}$).
\checkPM{The most significant difference between these two approaches
  resides in the fact that, when programming using the OpenMP
  fork-join model, the developer needs to identify which regions of
  the program will be executed in parallel.  Nevertheless, the
  overlapping of computation and communication phases can become
  challenging. On the other hand, in the OmpSs-2 tasking
  model\checkPMB{,} the parallelism is implicitly created from the
  beginning of the execution.  As a consequence, the developer needs
  to identify the tasks that will be executed asynchronously and their
  input/output dependencies to leverage its data-flow execution
  model. The tasking model requires more effort from the programmer
  but enables the transparent overlapping of computation and
  communication phases with the help of the TAMPI
  library~\cite{Perez2017,Sala2019}.}

\checkPM{Finally,} all the extensions that we made to the HPCCG
open-source code have been published at Code
Ocean~\citep{martinez2022} under the name HLAM (hybrid linear algebra
methods) to make them available to future readers.  The next
\checkPM{two} sections describe the hybrid implementations of the
CG-NB and the \emph{relaxed}, symmetric Gauss--Seidel methods within
HLAM.  The \checkPM{other algorithms} undergo a similar procedure and,
for the sake of conciseness, are omitted in this work.

\subsection{Parallelisation of the CG-NB algorithm}
\label{sec:parallelisation-CG-NB}
We follow the HDOT strategy that we recently
published~\cite{Ciesko2020} with the aim of facilitating the
programming of hybrid, task-based applications.  We define ``domains''
as the result of applying an MPI \emph{explicit} decomposition: a
spatial grid in this particular case.  It is explicit in the sense
that the programmer needs to design the application from scratch with
a very specific decomposition that is quite difficult to modify
afterwards because it requires major code refactoring.  Next, HDOT
defines a deeper hierarchy of ``subdomains'' to which one can assign
individual tasks that can be executed in parallel via appropriate
OpenMP or OmpSs-2 clauses with their corresponding read (\texttt{in})
and write (\texttt{out}) data dependencies.  This introduces
\emph{minimal} refactoring in MPI-only applications such as HPCCG thus
leaving the original MPI decomposition intact.  Finally, in order to
intertwine the two parallel programming models, e.g.\ MPI-OMP and
MPI-OSS, the communication tasks rely upon the TAMPI
library~\cite{Sala2019} which permits to overlap them with other
computation tasks transparently to the programmer.  It is worth
mentioning that the HDOT strategy targets truly hybrid applications
with zero sequential parts and zero implicit (fork-join) barriers.
Hence, we expect such hybrid applications to outperform their MPI-only
counterparts in most scenarios.

Code~\ref{code:CG-NB} shows two intermediate tasks (\texttt{Tk 1} and
\texttt{Tk 2} as described by Algorithm~\ref{alg:CG-NB}) inside the
main loop of the CG-NB method as well as the point-to-point MPI
communication of vector ${\bf r}$ (line 1) before the SpMV (line 12)
and the collective communication of $\alpha_d$ (line 24).  In order to
assign tasks to subdomains it is necessary to divide the iterative
space, that is \texttt{nrow}, which represents the actual length of
the involved arrays after the explicit decomposition of MPI.  This is
done with an external loop (line 6) and appropriate starting indexes
and block sizes (line 7).  The total number of tasks is provided by
the user at runtime and determines the granularity.  As a rule of
thumb, there should be enough tasks to feed all the cores; however,
having too many of them will cause overheads and ultimately impact the
parallel performance.  Moving forward, the task computing the SpMV
(lines 9--12 in Code~\ref{code:CG-NB}) needs a \emph{multidata} read
dependency on ${\bf r}$ and a \emph{regular} write dependency on a
region of \texttt{Ar} that stores the sparse matrix--vector product
${\bf A} \cdot {\bf r}$ (line 11).  Indeed, the SpMV kernel has the
particularity of accessing noncontiguous elements of \texttt{r} thus
yielding an irregular data access pattern, which we precalculate and
store in variables \texttt{depMVidx} and \texttt{depMVsize} at the
beginning of the program for later reuse.  The next task (lines
14--21) performs altogether two vector updates and one scalar product
using regular data dependencies.  It is worth mentioning the use of
the macro \texttt{ISODD}, which allows us to carry out independent MPI
communications along two consecutive iterations by exploiting
different communicators.  In practice, this prevents any possibility
of MPI deadlocks associated with different iterations of the main
loop.

Special attention must be paid to local reductions via tasks when
calculating, for instance, $\alpha_{\rm d}$ at line 21 of
Code~\ref{code:CG-NB}.  Contrarily to the MPI-only or MPI-OMP$_{\rm
  fj}$ implementations, the task execution order is not guaranteed
between consecutive iterations or two complete executions of the CG-NB
algorithm.  As a result, floating\checkPMB{-}point rounding errors can
accumulate and affect the convergence rate of the entire algorithm.
Whilst this does not constitute an issue for the CG methods, it does
have devastating consequences for the BiCGStab algorithm and its
derivatives.  It is well known that BiCGStab is prone to divergence
when the residual \checkPM{$\sqrt{\alpha_{n,j+1}}$ (see line 10 of
  Algorithm~\ref{alg:BiCGStab-B1})} gets closer to zero.  Tasks do
aggravate this situation and, as a result, each BiCGStab execution is
susceptible to reaching convergence a \checkPM{handful number of
  iterations later} \checkPMB{(convergence is typically attained after
  a few tenths of iterations or less)}.  To prevent this from
happening, we employ a restart procedure (lines 13--15 of
Algorithm~\ref{alg:BiCGStab-B1}) that triggers whenever the residual
value reaches a given threshold (typically $\varepsilon = 10^{-5}$,
\checkPMB{although it may be problem-specific and require additional
  tunning}).  Restarting not only reduces the total number of
iterations to reach convergence for all the parallel implementations
tested in this work, but it also eliminates the accumulative rounding
errors introduced by the task execution order.  \checkPMB{Note,
  however, that this nondeterministic execution order may not be
  desirable for high fidelity simulations where instantaneous fields
  need to be accurately reproduced.}

The function \texttt{exchange\_externals} performing the
point-to-point communication is shown in Code~\ref{code:MPI}.
Neighbouring data is received close to \checkPM{the} end of buffer
\texttt{x} (line 8) to later be used by the SpMV kernel.  As for
sending data, it is a bit more complicated due to the irregular access
pattern caused by the explicit MPI decomposition.  Therefore, the use
of multidata dependencies becomes necessary again.  Note that in one
single task data are copied, arranged contiguously in a temporary
buffer, and sent to the corresponding neighbour.  These two
communication tasks (lines 6--9 and 17--23) are transparently
scheduled and executed along with computation tasks with the aid of
the TAMPI library.  The \texttt{TAMPI\_IWAIT} specialised nonblocking
function converts the MPI synchronicity into a regular task data
dependency handled automatically by the OmpSs-2 runtime (the same
holds true with OpenMP tasks and the OpenMP runtime).

Hybrid parallelism can also be achieved by combining MPI with OpenMP
and its fork-join model, which today constitutes the most common
archetype of hybrid programming.  Code~\ref{code:SpMV} shows how this
model is supported in the SpMV function in such a way that still
remains compatible with the task-based implementation described above.
Within each subdomain of length \texttt{size}, the iteration space is
further subdivided in blocks of size \texttt{bs} and assigned to
different OpenMP threads, see lines 1--11.  In practice, however, when
the fork-join model is enabled the number of subdomains is reduced to
one as there are no tasks present.  The same strategy is adopted in
other functions (\texttt{axpby}, \texttt{dot}, etc.).  The amount of
work distributed to each thread is carried out by the function
\texttt{split} at line 2 of Code~\ref{code:SpMV}.  This subroutine
attempts to align partitions to boundaries compatible with the
specified SIMD vector length whenever possible (see lines 16--17).
This function is also employed to determine the best number of
subdomains (tasks) while seeking optimal vectorisation properties.  As
it is the case \checkPM{when using, for example,} Intel MKL parallel
functions, fork-join parallelism remains \emph{internal} to the kernel
and, therefore, switching from one kernel to another yields an
\emph{implicit barrier} that is not present in the task-based
implementation \checkPM{(although the use of \texttt{nowait} clauses
  may help in eliminating such barriers)}.  Finally, this fork-join
implementation cannot handle MPI collectives via tasks and, therefore,
it reuses MPI synchronous calls (\texttt{MPI\_Wait} and
\texttt{MPI\_Allreduce}) originally implemented in HPCCG.

\subsection{Parallelisation of the relaxed Gauss--Seidel algorithm}
\label{sec:parallelisation-relaxed-Gauss-Seidel}
In the particular case of the symmetric Gauss--Seidel solver, the most
adopted task-based strategy consists of assigning colours to
subdomains or tasks.  This brings new opportunities for
parallelisation by \emph{assuming} that each colour can be calculated
\emph{independently} so that data dependencies only affect tasks
belonging to the same colour.  However, this assumption does not
guarantee the exact same execution order of the original algorithm,
which is essentially sequential.  As a result, the parallel algorithm
can exhibit a different convergence behaviour.  Our \emph{standard}
task-based parallelisation of the GS algorithm supports multicolouring
and colour rotation between iterations but, for practical reasons, a
red--black colour scheme without rotation is adopted herein.  Indeed,
for the sparse matrices used in this work, this basic strategy gives
us the best performance and no further advantages are observed with
the addition of more colours.  \checkPM{This can be explained by the
  fact that both HLAM and HPCCG codes work on structured meshes.}

We decide to go one step further in the parallelisation of the GS
method by assigning a new set of task clauses that do not complain
with the red--black colouring scheme but, on the other hand, do reduce
the complexity of task data dependencies.
Code~\ref{code:relaxed-Gauss-Seidel} shows the computation tasks
inside the main loop of this \emph{relaxed} implementation.  The
forward and backward sweeps contain a regular write dependency on
vector ${\bf x}$ instead of the expected multidependency that is
present in the bicoloured task-based implementation.  Each sweep is
self-contained in a different loop (lines 8--12 and 15--20) to
guarantee that tasks associated with forward sweeps over a particular
subdomain are computed before the backward sweep.  We introduce an
extra level of task synchronicity when initialising the local residual
at lines 1--6.  This is to avoid any computation overlaps between
different iterations.  The task annotation shown in
Code~\ref{code:relaxed-Gauss-Seidel} guarantees numerical convergence
because the data races that are created mimic the Gauss--Seidel
behaviour in which previously calculated data are being continuously
reused within the current iteration.  Not having data races will imply
a slowdown in convergence for this relaxed implementation, making it
similar to that of the Jacobi method.

\section{Results and discussion}
\label{sec:results-and-discussion}
This section evaluates the CG, BiCGStab, symmetric Gauss--Seidel and
Jacobi original methods as well as the proposed alternative versions.
Similarly, different parallelisation strategies are compared against
each other and analysed in detail.

\subsection{Numerical setup}
We carry out the numerical experiments on the MareNostrum 4
supercomputer, located at the Barcelona Supercomputer Center (BSC).
Each computational node is equipped with two Intel Xeon Platinum 8160
CPUs composed of 24 physical cores (hyperthreading disabled) running
at a fixed frequency of $\mathrm{2.1 \, GHz}$.  The L3 cache memory of
this CPU is $\mathrm{33 \, MiB}$.  We pay special attention to this
number and verify that our memory-bound applications are actually
limited by the system main memory (RAM).  \checkPM{To ensure this,}
each process of a MPI-only application works on a grid size of $128
\times 128 \times 128$ elements and, in the case of a hybrid
application (1 MPI process per socket), this quantity is increased by
a factor of 24 ($128 \times 128 \times 3072$, note that
\checkPMB{HPCCG, and thus HLAM, only distribute} points along the
  last dimension).  This implies 384 MiB per array (stored in double
  precision), which is one order of magnitude larger than the L3
  cache.  For a fair comparison between different methods and parallel
  implementations, results presented in the same graph are run within
  the same job script, thus using the exact same computational
  resources, and repeated up to ten times in order to extract relevant
  statistics.  A version of Clang 15 modified by
  BSC\footnote{https://github.com/bsc-pm/ompss-2-releases} constitutes
  the predetermined compiler: it allows us to quickly switch between
  OpenMP and OmpSs-2 runtimes.  \checkPM{The} source code is compiled
  with the usual flags \texttt{-march=native} and \texttt{-Ofast}
  together with support for 512-bit SIMD vector instructions and
  memory alignment to 2 MiB transparent huge pages.  \checkPM{Finally,
    Intel's MPI library 2018.4 is used for all the experiments carried
    out on MareNostrum 4.}

\checkPM{The sparse linear system to be solved is the standard one
  proposed by the HPCG benchmark~\citep{Dongarra2016} and arises from
  the finite discretisation of a centred stencil on a
  three-dimensional hexahedral mesh.  The r.h.s.\ vector $\mathbf{b}$
  is defined analytically for the exact solution $\mathbf{x = 1}$.
  Next, the numerical method is initialised to $\mathbf{x = 0}$ and is
  let to evolve with the convergence criteria set to $\epsilon =
  10^{-6}$ together with a restarting threshold value of $\varepsilon
  = 10^{-5}$ for the BiCGStab methods.  The required number of
  iterations to reach convergence depends strongly on the numerical
  method and the degree of sparsity of the system matrix and, to a
  much lesser extent, the parallel implementation.  When using one
  compute node of MareNostrum 4 and the 7-point centred stencil, the
  required iterations are: 8 (BiCGStab), 12 (CG), 9 (Gauss--Seidel),
  and 18 (Jacobi); similarly, the iterations associated with the
  27-point stencil are: 45 (BiCGStab), 72 (CG), 142 (Gauss--Seidel),
  and 515 (Jacobi).}

\begin{figure}[H]
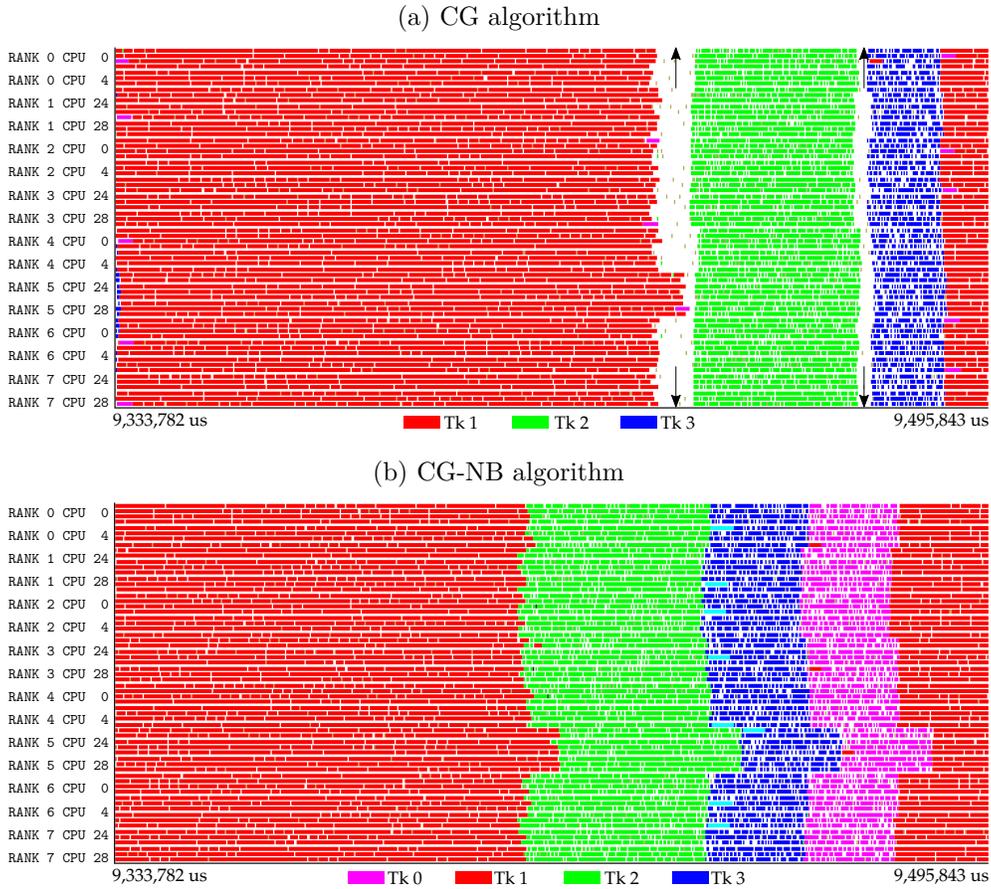

  \centering
  \subfloat[CG algorithm]{
    \includegraphics[width=0.95\columnwidth]{\figSrc{trace-CG}}
  }
\\
  \subfloat[CG-NB algorithm]{
    \includegraphics[width=0.95\columnwidth]{\figSrc{trace-CG-NB}}
  }
  \caption{Paraver traces associated with the (a) classical CG and (b)
    nonblocking CG (CG-NB) hybrid methods implemented via
    MPI-OSS$_{\rm t}$ using 8 MPI ranks and 8 cores per rank.  Events
    highlighted in different colours correspond to OmpSs-2 tasks.
    Arrows indicate blocking barriers due to MPI collective
    communication.}
  \label{fig:traces-CG}
\end{figure}

\subsection{Weak scalability of Krylov subspace methods}
We compare the CG and BiCGStab methods using several parallelisation
strategies.  For task-based implementations, we first need to identify
the optimal task granularity\checkPM{, which mainly depends on the
  task dependency graph associated with the SpMV operation.  This
  process, omitted here for the sake of conciseness, can be automated
  by executing the application binary with an increasing number of
  tasks multiple of the available resources (24 cores per socket on
  MareNostrum 4). The optimal task granularity} turns out to be around
800 tasks and 1500 tasks depending on the matrix sparsity pattern,
that is 7- and 27-point stencils, respectively.  It is worth
mentioning that there is a fair interval around these numbers where
the performance still remains optimal.  The granularity is mainly
determined by the most computationally intensive kernel, i.e.\ SpMV,
and for this reason the optimal granularity is quite similar among
different Krylov methods.  The same applies to the Jacobi and
symmetric Gauss--Seidel methods, where the algorithm access pattern
resembles quite a lot that of the SpMV with additional operations over
the diagonal elements of the matrix.

Before comparing different methods and implementations against each
other, it is worth discussing the improvements brought by the proposed
algorithms.  Figure~\ref{fig:traces-CG} shows two Paraver traces
corresponding to the original and nonblocking CG methods.  They are
obtained after executing the respective hybrid implementations making
use of MPI and OmpSs-2 tasks under the exact same computational
resources: 8 MPI ranks and 8 cores per rank in order to improve the
readability of the trace.  The time interval, which is the same for
both traces, is sufficiently large to cover one entire loop iteration
approximately.  For the classical version,
Fig.~\ref{fig:traces-CG}(a), the presence of two synchronisation
barriers (indicated by arrows) due to collective MPI communication is
evident.  For the nonblocking method, Fig.~\ref{fig:traces-CG}(b),
coloured events correspond to different tasks as described in
Algorithm~\ref{alg:CG-NB}.  The same labels and colours have been
retained on both figures although there is not a straight one-to-one
mapping of tasks between these two methods.  From the results shown
above, it can be stated that the proposed CG-NB algorithm, when
executed via the MPI-OSS$_{\rm t}$ hybrid model, can effectively
suppress the two blocking barriers present in the original method.
The same can be said for the BiCGStab-B1 method, where the hybrid
implementation based on tasks does eliminate two of the three global
barriers.  We purposely omit the corresponding traces for the sake of
brevity.

\begin{figure}[H]
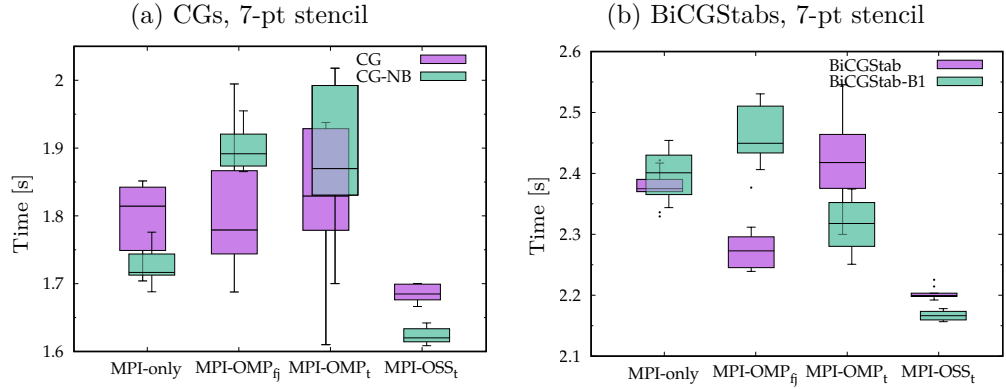

  \centering
  \subfloat[CGs, 7-pt stencil]{
    \martinezFigLabelTwo{0.43\columnwidth}{\figSrc{CG-parImp-time-16node}}
    {Time}{[s]}
  }
  \subfloat[BiCGStabs, 7-pt stencil]{
    \martinezFigLabelTwo{0.43\columnwidth}{\figSrc{BiCGStab-parImp-time-16node}}
    {Time}{[s]}
  }
  \caption{Execution time comparison between several parallel
    implementations of the classical and the proposed variants of the
    (a) CG and (b) BiCGStab methods on 16 compute nodes of MareNostrum
    4 using the 7-point stencil pattern.  \checkPM{Standard box and
      whisker plots.}}
  \label{fig:CG-BiCGStab-time-16node}
\end{figure}

The objective of
\checkPM{Figs.~\ref{fig:CG-BiCGStab-time-16node}(a)--(b) (box and
  whisker plots)} is to show the execution time variability of various
parallel implementations of KVMs using 16 compute nodes and the
sparsest coefficient matrix associated with a 7-point spatial stencil.
Starting with the classical CG version,
Fig.~\ref{fig:CG-BiCGStab-time-16node}(a), the MPI-only implementation
and the two hybrid ones based on OpenMP report similar median times
with significant variations between executions.  On the other hand,
the task-based OmpSs-2 implementation yields a significant
improvement, both in terms of execution time and variability, being
the median execution time 7.7\% lower with respect to the MPI-only
implementation.  When switching to the nonblocking CG method,
MPI-OMP$_{\rm fj}$ and MPI-OMP$_{\rm t}$ versions do report larger
times due to the extra number of \checkPM{memory accesses (see
  Section~\ref{sec:description-algorithms})}; to our surprise, this is
not the case for the MPI-only implementation.  The nonblocking,
OmpSs-2 version shows an \emph{additional} improvement of 4\% over its
classical counterpart.  Moving on to the original BiCGStab method, see
Fig.~\ref{fig:CG-BiCGStab-time-16node}(b), the OpenMP fork-join and
OmpSs-2 task-based implementations offer speedups of 4.5\% and 12\%
over the MPI-only version, respectively.  The task-based
implementations of the BiCGStab-B1 algorithm can suppress two of the
three synchronous points per iteration by overlapping computation with
communication tasks via the TAMPI library.  This leads to relative
improvements of 4.3\% and 1.5\% for the OpenMP and OmpSs-2 runtimes,
respectively.  As expected, they remain rather modest compared to
those seen for CG since there still remains one unavoidable
synchronisation at line 3 of Algorithm~\ref{alg:BiCGStab-B1}.
Finally, it is worth noting the remarkable reduction in execution
variability achieved by the tight integration of MPI and OmpSs-2
programming models via the TAMPI library.  The remainder of this work
focuses on the OmpSs-2 task-based version due to its superiority over
OpenMP tasks and evaluates both the weak and strong scalabilities of
four algorithms and three parallel implementations over a broad range
of compute nodes using two sparsity degrees for the coefficient
matrix.

\begin{figure}[H]
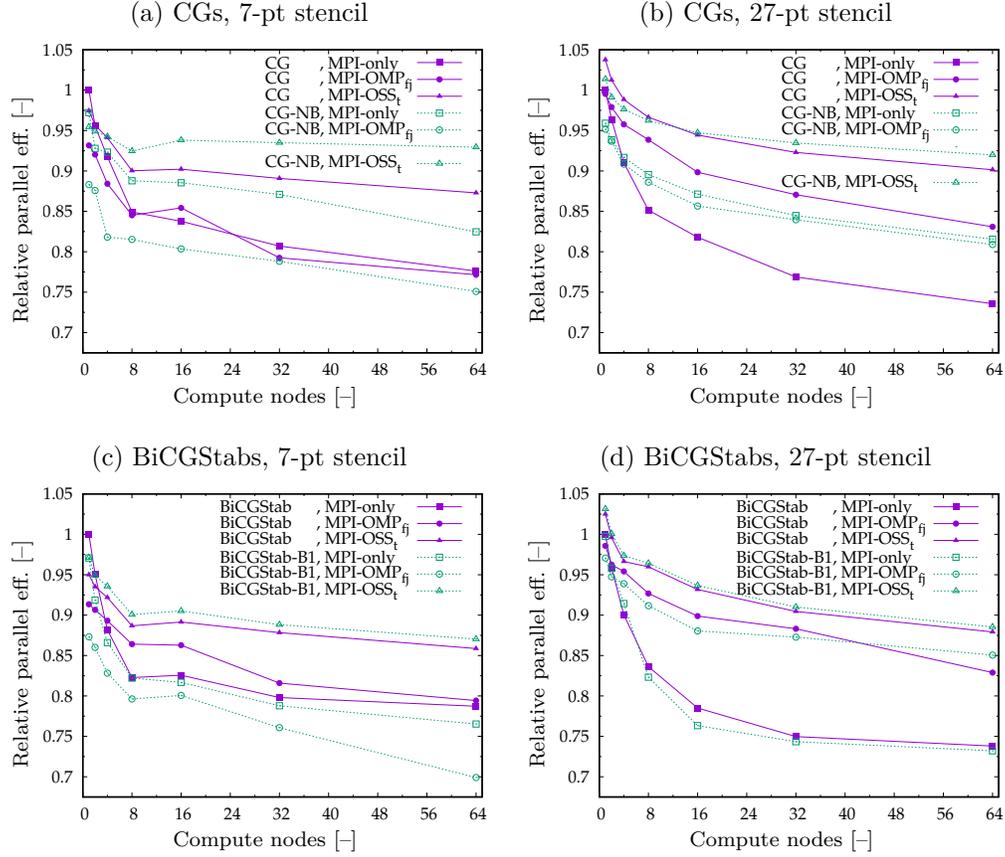

  \centering
  \subfloat[CGs, 7-pt stencil]{
    \martinezFigLabel{0.43\columnwidth}{\figSrc{CG-weakScal-7}}
    {Compute nodes}{[--]}
    {Relative parallel eff.}{[--]}
  }
  \subfloat[CGs, 27-pt stencil]{
    \martinezFigLabel{0.43\columnwidth}{\figSrc{CG-weakScal-27}}
    {Compute nodes}{[--]}
    {Relative parallel eff.}{[--]}
  }
\\
\subfloat[BiCGStabs, 7-pt stencil]{
    \martinezFigLabel{0.43\columnwidth}{\figSrc{BiCGStab-weakScal-7}}
    {Compute nodes}{[--]}
    {Relative parallel eff.}{[--]}
  }
  \subfloat[BiCGStabs, 27-pt stencil]{
    \martinezFigLabel{0.43\columnwidth}{\figSrc{BiCGStab-weakScal-27}}
    {Compute nodes}{[--]}
    {Relative parallel eff.}{[--]}
  }
  \caption{Relative parallel efficiencies associated with the weak
    scalability analysis of the CG and BiCGStab methods with two
    sparsity patterns using up to 64 compute nodes of MareNostrum 4.
    Median reference times correspond to the MPI-only, classical
    implementations on one compute node (first \checkPM{filled}
    square): (a) 1.52s, (b) 19.35s, (c) 1.96s and (d) 23.76s.  Each
    point represents the median value of 10 executions.}
  \label{fig:CG-BiCGStab-weakScal}
\end{figure}

Figure~\ref{fig:CG-BiCGStab-weakScal} shows the results derived from a
weak scalability analysis of the CG and BiCGStab methods using several
parallel implementations and sparsity patterns.  To facilitate the
comparison, times will always be normalised by the MPI-only, classical
version of each algorithm executed on one compute node, which assigns
the value of 1 to the first \checkPM{filled} square point of each
graph.  Consequently, these figures do report \emph{relative} parallel
efficiencies and, for this reason, some values may exceed unity.
Absolute execution time values can be easily retrieved from graphs
with the specified reference times.  Beginning with the classical CG
methods, Figs.~\ref{fig:CG-BiCGStab-weakScal}(a)-(b), the OmpSs-2
version turns out to be overall much faster than its MPI-only
counterpart except when employing one node and the 7-point stencil.
By switching from the classical algorithm to the nonblocking one, this
task-based implementation becomes 19.7\% and 25\% faster than the
MPI-only version for the 7- and 27-point stencils, respectively, when
using 64 compute nodes.  The fork-join hybrid implementation yields
mixed results, being only faster than the MPI-only version when
considering the 27-point stencil.  As it was the case of
Fig.~\ref{fig:CG-BiCGStab-time-16node}, the proposed nonblocking
algorithm only benefits the MPI-only and task-based versions and is
counterproductive with fork-join parallelism.  Regarding the
BiCGStab methods, Figs.~\ref{fig:CG-BiCGStab-weakScal}(c)-(d), results
are generally similar to those observed for CG with minor
discrepancies.  For instance, only the task-based implementations take
advantage of the BiCGStab-B1 algorithm by a small, but consistent
margin.  Therefore, they remain faster than the classical algorithm by
1.4\% and 0.7\% for the 7- and 27-point stencils, respectively, when
using the largest set of compute nodes.  As previously mentioned, such
small percentages are expected because of the remaining global
barrier.  \checkPM{For} this same node configuration, the BiCGStab
task-based version is, depending on the sparsity pattern, 10.6\% and
20\% faster than the MPI-only classical implementation and overall
more efficient than the fork-join hybrid version.

\checkPMB{Another} aspect worth mentioning about
Fig.~\ref{fig:CG-BiCGStab-weakScal} is the fact that the MPI-only
implementations of the CG and BiCGStab methods do not seem to benefit
from the increase in computational intensity caused by the 27-point
stencil.  Indeed, the relative parallel efficiencies are equal or
worse than those obtained with the 7-point stencil, contrarily to what
the shared-memory strategies report.  In any case, we do not expect
large discrepancies when switching stencils since the problem size is
large enough to guarantee that each kernel within the application
remains memory bound.

\checkPMB{A final note must be added with regard to the quick
  degradation of MPI-only implementations with an increasing number of
  compute nodes.  For instance, Fig.~\ref{fig:CG-BiCGStab-weakScal}(a)
  reveals that the relative parallel efficiency of the CG method drops
  15\% with eight compute nodes (384 MPI processes).  Such a loss in
  performance is mostly related to the computation of the two dot
  products and, more specifically, to the corresponding global
  reductions due to the increasing number of internode communications.
  Although \texttt{MPI\_Allreduce} synthetic benchmarks report typical
  latencies within the order of $\mathrm{10^{-5} \, s}$ for small
  message sizes on MareNostrum 4, we can measure latencies of about
  $\mathrm{10^{-3} \, s}$ on average for the CG method.  The two MPI
  collective communications shown in Fig.~\ref{fig:traces-CG}(a) are
  good examples of these time intervals.  Contrarily to a synthetic
  benchmark, a complex application like CG is more sensitive to system
  load imbalances and noise affecting computations between two
  consecutive collectives.  As a result, the \emph{effective}
  communication time spent in global communications can be up to two
  orders of magnitude larger than the \emph{minimum} latency retrieved
  from a simple benchmark.  These overheads accumulate iteration after
  iteration and add up to the total execution time.  MPI-only
  applications tend to suffer more from load-balancing issues and this
  situation aggravates with the increasing core count of new CPUs.}

\begin{figure}[H]
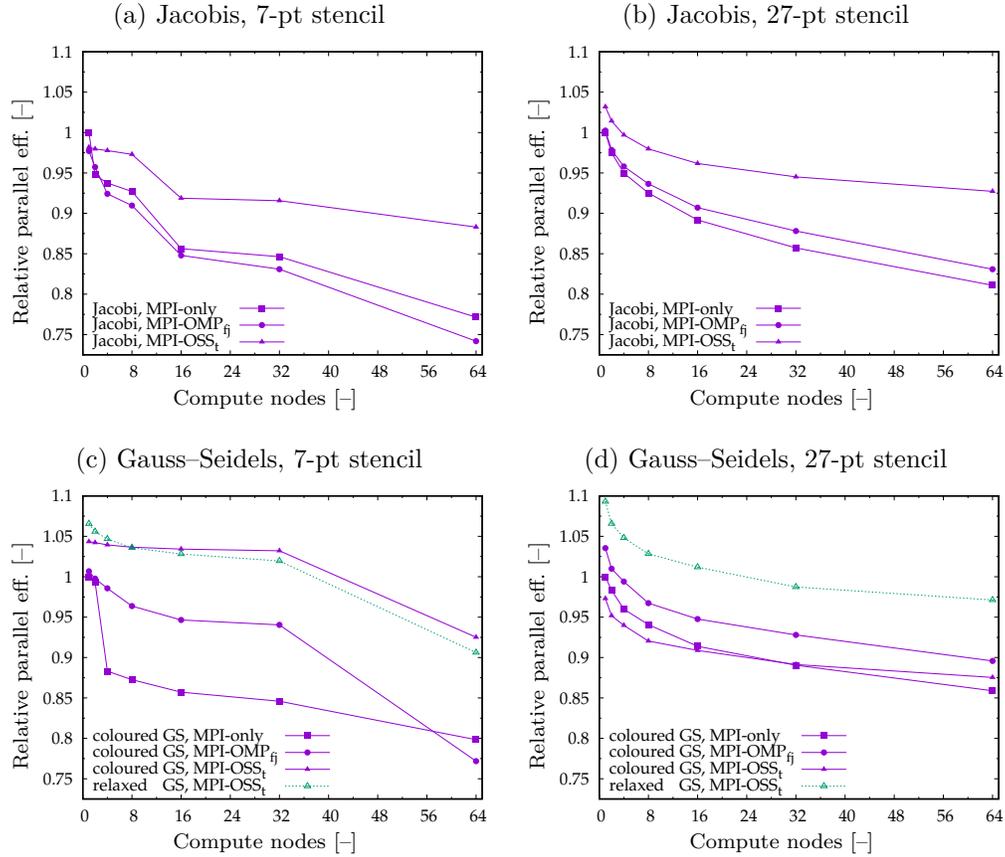

  \centering
  \subfloat[Jacobis, 7-pt stencil]{
    \martinezFigLabel{0.43\columnwidth}{\figSrc{Jacobi-weakScal-7}}
    {Compute nodes}{[--]}
    {Relative parallel eff.}{[--]}
  }
  \subfloat[Jacobis, 27-pt stencil]{
    \martinezFigLabel{0.43\columnwidth}{\figSrc{Jacobi-weakScal-27}}
    {Compute nodes}{[--]}
    {Relative parallel eff.}{[--]}
  }
\\
  \subfloat[Gauss--Seidels, 7-pt stencil]{
    \martinezFigLabel{0.43\columnwidth}{\figSrc{GaussSeidel-weakScal-7}}
    {Compute nodes}{[--]}
    {Relative parallel eff.}{[--]}
  }
  \subfloat[Gauss--Seidels, 27-pt stencil]{
    \martinezFigLabel{0.43\columnwidth}{\figSrc{GaussSeidel-weakScal-27}}
    {Compute nodes}{[--]}
    {Relative parallel eff.}{[--]}
  }
  \caption{Relative parallel efficiencies associated with the weak
    scalability analysis of the Jacobi and symmetric Gauss--Seidel
    methods with two sparsity patterns using up to 64 compute nodes of
    MareNostrum 4.  Median reference times correspond to the MPI-only
    reference implementations on one compute node (first
    \checkPM{filled} square): (a) 1.40s, \checkPM{(b)} 113.91s,
    \checkPM{(c)} 1.31s and (d) 61.65s.  Each point represents the
    median value of 10 executions.}
  \label{fig:Jacobi-GaussSeidel-weakScal}
\end{figure}

\subsection{Weak scalability of Jacobi and symmetric Gauss-Seidel methods}
We carry out the same scalability study for the Jacobi and symmetric
Gauss--Seidel methods and report the relative parallel efficiencies in
Fig.~\ref{fig:Jacobi-GaussSeidel-weakScal}.  A similar legend and
nondimensionalisation of results are employed to help facilitating the
comparison with previous Krylov methods.  Jacobi is the most
straightforward algorithm and one unique kernel is written using three
different parallel implementations.  While the MPI-only and
MPI-OMP$_{\rm fj}$ methods do report more or less similar execution
times, see Figs.~\ref{fig:Jacobi-GaussSeidel-weakScal}(a)-(b), the
MPI-OSS$_{\rm t}$ is able to bring more performance to the table.
With 64 compute nodes, MPI-OSS$_{\rm t}$ offers a 14.4\% and 14.3\%
speedup over the MPI-only version employing the 7- and 27-point
stencils, respectively.  In the case of the symmetric Gauss--Seidel
algorithm, we show the typical task-coloured implementation as well as
the data-relaxed version implemented with tasks and compare them
against the MPI-only and fork-join hybrid alternatives.  On the one
hand, the fork-join approach seems to have a competitive advantage
over pure MPI with the exception of the last point of
Fig.~\ref{fig:Jacobi-GaussSeidel-weakScal}(c).  On the other hand, the
two task-based implementations can behave quite differently.  When
using the sparsest matrix, the red--black colouring strategy can offer
a competitive advantage on large core counts, but not for a large
margin.  However, when the matrix becomes less sparse,
Fig.~\ref{fig:Jacobi-GaussSeidel-weakScal}(d), the task-based approach
based on two colours may result in terrible performance.  Contrarily,
the relaxed version excels in this particular case.  These remarkable
differences in efficiency can be easily explained by the number of
iterations that each parallel implementation of the GS necessitates to
reach numerical convergence.  While the MPI-only version convergences
in 157 iterations, the bicoloured task-based GS needs 166 and the
relaxed version finishes at 150; for reference, the fork-join
implementation takes 152 iterations.  One can reduce this number of
iterations of the coloured version by simply coarsening the task
granularity.  However, we do not recommend this strategy since
combining the symmetric Gauss--Seidel method as a preconditioner with a
Krylov algorithm, and using different granularities for each
technique, will certainly derive in bottlenecks.  Finally, taking into
account the best task-based GS implementation on 64 nodes, the
obtained speedups are 15.9\% and 13.1\% with respect to the MPI-only
version using the 7- and 27-point stencils, respectively.

\subsection{Strong scalability of hybrid iterative methods}
The last battery of results correspond to strong scalability tests in
which the problem size of $128 \times 128 \times 6144$ elements does
not longer increase with the number of computational resources.  In
order to avoid excessive repetition, for each parallel implementation
we only consider the overall best performing algorithm: classical or
proposed variant.  Regarding task-based hybrid implementations, we use
the optimal task granularity, which we calculate in advance.  For
convenience, the next figures make use of the same legend and
nondimensionalisation specified for the weak scalability graphs.

\begin{figure}[H]
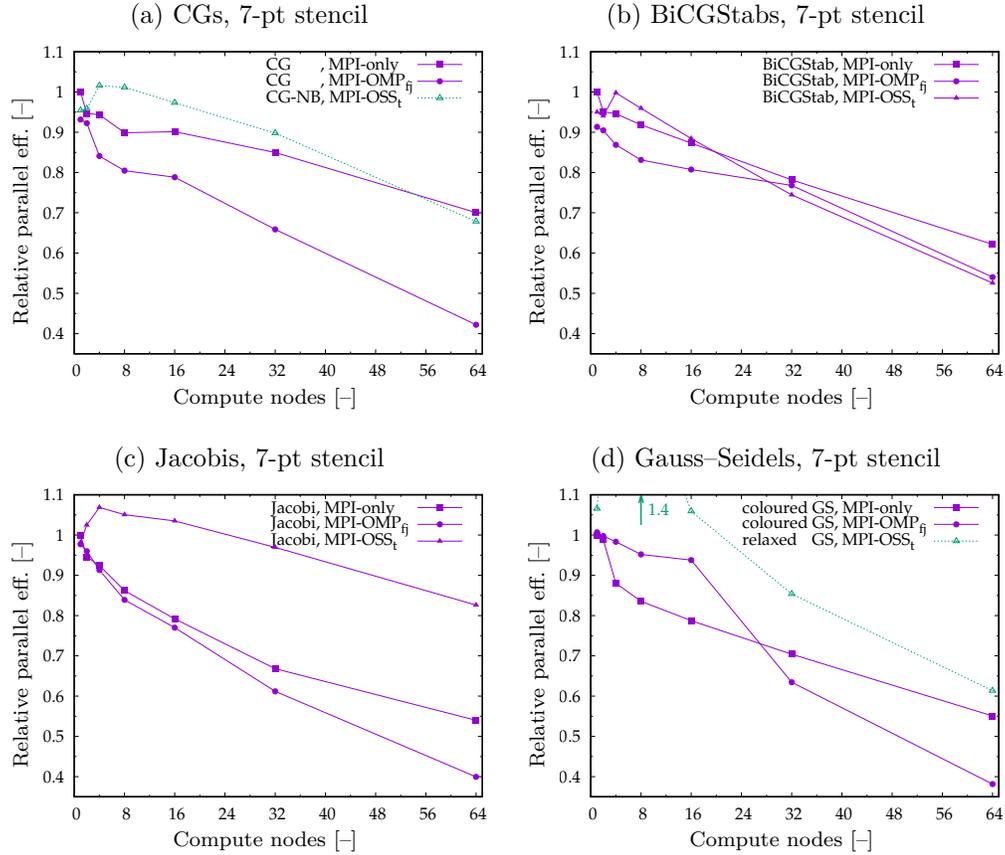

  \centering
  \subfloat[CGs, 7-pt stencil]{
    \martinezFigLabel{0.43\columnwidth}{\figSrc{CG-strongScal-7}}
    {Compute nodes}{[--]}
    {Relative parallel eff.}{[--]}
  }
  \subfloat[BiCGStabs, 7-pt stencil]{
    \martinezFigLabel{0.43\columnwidth}{\figSrc{BiCGStab-strongScal-7}}
    {Compute nodes}{[--]}
    {Relative parallel eff.}{[--]}
  }
\\
  \subfloat[Jacobis, 7-pt stencil]{
    \martinezFigLabel{0.43\columnwidth}{\figSrc{Jacobi-strongScal-7}}
    {Compute nodes}{[--]}
    {Relative parallel eff.}{[--]}
  }
  \subfloat[Gauss--Seidels, 7-pt stencil]{
    \martinezFigLabel{0.43\columnwidth}{\figSrc{GaussSeidel-strongScal-7}}
    {Compute nodes}{[--]}
    {Relative parallel eff.}{[--]}
  }
  \caption{Relative parallel efficiencies associated with the strong
    scalability analysis of four iterative methods using the 7-point
    stencil pattern.  Same legend and nondimensionalisation as in
    Figs.~\ref{fig:CG-BiCGStab-weakScal}(a),(c) and
    Figs.~\ref{fig:Jacobi-GaussSeidel-weakScal}(a),(c).}
  \label{fig:strongScal-7pt}
\end{figure}

Figure~\ref{fig:strongScal-7pt} shows the strong scalability results
of the four iterative methods considered in this work using the
sparsest matrix derived from a 7-point stencil.  As expected, when the
problem size begins to fit within the L3 cache, the computational
advantage of tasks vanishes due to data locality issues.  In such
scenarios, either the pure MPI approach or the fork-join
implementation will exploit data locality better.  As the number of
computational resources keeps growing, MPI-OSS$_{\rm t}$
implementations of the two KVM methods,
Figs.~\ref{fig:strongScal-7pt}(a)-(b), report quite similar or even
subpar performances with respect to their corresponding MPI-only
versions.  The iterative methods of Jacobi and, in particular, the
relaxed Gauss--Seidel do exhibit superscalability when executed via
OmpSs-2 tasks, see Figs.~\ref{fig:strongScal-7pt}(c)-(d).  In general,
the fork-join hybrid implementations do not seem to represent a real
improvement over MPI-only and often result in the slowest parallel
implementation.  Finally, note that the proposed BiCGStab-B1 method is
not shown in Fig.~\ref{fig:strongScal-7pt}(b) since it yields overall
worse results than the classical algorithm for strong scalability
scenarios.  The same situation is observed when using the less sparse
matrix derived from the 27-point spatial stencil.

\begin{figure}[H]
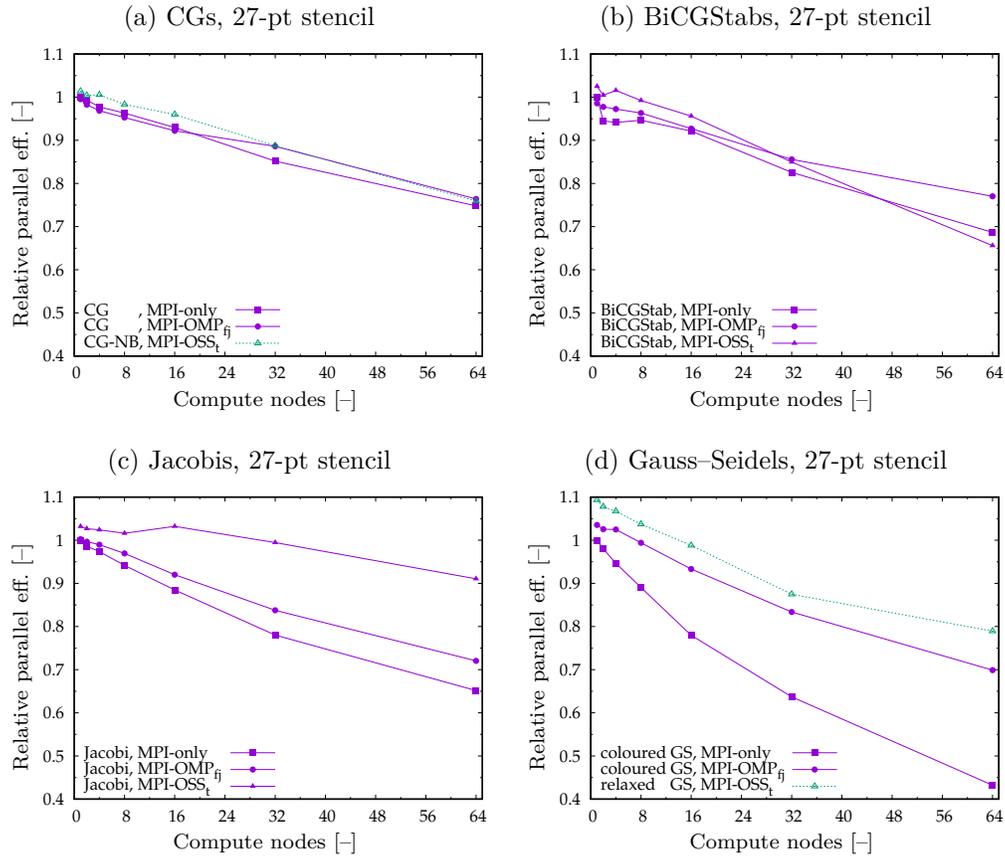

  \centering
  \subfloat[CGs, 27-pt stencil]{
    \martinezFigLabel{0.43\columnwidth}{\figSrc{CG-strongScal-27}}
    {Compute nodes}{[--]}
    {Relative parallel eff.}{[--]}
  }
  \subfloat[BiCGStabs, 27-pt stencil]{
    \martinezFigLabel{0.43\columnwidth}{\figSrc{BiCGStab-strongScal-27}}
    {Compute nodes}{[--]}
    {Relative parallel eff.}{[--]}
  }
\\
  \subfloat[Jacobis, 27-pt stencil]{
    \martinezFigLabel{0.43\columnwidth}{\figSrc{Jacobi-strongScal-27}}
    {Compute nodes}{[--]}
    {Relative parallel eff.}{[--]}
  }
  \subfloat[Gauss--Seidels, 27-pt stencil]{
    \martinezFigLabel{0.43\columnwidth}{\figSrc{GaussSeidel-strongScal-27}}
    {Compute nodes}{[--]}
    {Relative parallel eff.}{[--]}
  }
  \caption{Relative parallel efficiencies associated with the strong
    scalability analysis of four iterative methods using the 27-point
    stencil pattern.  Same legend and nondimensionalisation as in
    Figs.~\ref{fig:CG-BiCGStab-weakScal}(b),(d) and
    Figs.~\ref{fig:Jacobi-GaussSeidel-weakScal}(b),(d).}
  \label{fig:strongScal-27pt}
\end{figure}

The last results from strong scalability benchmarks carried out with
the 27-point stencil are summarised in
Figure~\ref{fig:strongScal-27pt}.  When looking at the CG and BiCGStab
methods, the results are close to those obtained in
Fig.~\ref{fig:strongScal-7pt}: the task-based versions start with a
competitive advantage that cancels out progressively with the number
of computational resources.  Nonetheless, the three parallel
implementations report similar relative efficiencies and hence there
are not remarkable differences between them in terms of execution
time, see Figs.~\ref{fig:strongScal-27pt}(a)-(b).  The hybrid
implementation based on tasks reports the best efficiency figures for
the Jacobi and GS methods followed after by the fork-join model.  In
this particular case, the ubiquitous MPI-only parallelisation is at a
clear disadvantage against the two hybrid ones.

\section{Conclusions and future work}
\label{sec:conclusions-future-work}
In this work, we have presented the hybrid implementation \checkPM{on
  CPUs} of four iterative methods, namely: Jacobi, symmetric
Gauss--Seidel, conjugate gradient, and biconjugate gradient
\checkPM{stabilized}.  Alternative versions of these algorithms,
reducing the number of blocking barriers or completely eliminating
them, have also been proposed and compared against the original ones.
The hybridisation via OpenMP fork-join parallelism as well as via
OpenMP and OmpSs-2 tasks follows the HDOT methodology that we
described in a previous paper.  Specific code examples are described
in detail in this document whilst the entire HLAM source code has been
made publicly available at Code Ocean for reproducibility and also to
be used as a guide for hybrid programming and code implementation of
linear algebra iterative methods.

The weak scalability tests asseverated the superiority of task-based
hybrid implementations over the well-known MPI-only and fork-join
hybrid implementations.  In general, hybrid codes based on the popular
OpenMP fork-join paradigm have yielded mixed results, offering
sometimes similar or even subpar performance with respect to the
MPI-only implementation.  The nonblocking alternative versions
proposed in this document have also demonstrated to further improve
the \checkPM{overall} efficiency of the algorithms.  The battery of
strong scalability benchmarks has shown a competitive advantage of
hybrid parallel methods based on tasks for moderate computational
resources where data locality effects are not important.  While the
reported efficiencies remained quite similar for the three parallel
implementations of Krylov methods, adding tasks to Jacobi and
Gauss--Seidel offered a notorious speedup over MPI-only and
MPI-OMP$_{\rm fj}$ models.

This paper demonstrates the appropriateness of bringing hybrid
parallelism, in particular task-based parallelism, into numerical
libraries such as OpenFOAM \checkPM{and PETSc}, which extensively
exploit the numerical methods presented herein.  As a future work, we
plan to include these hybrid algorithms in a
\checkPM{\emph{minimalist}} version of this popular CFD library and
report the observed gains.  In addition to this, we are planning to
\checkPM{continue} our code developments over the popular HPCG
benchmark\checkPM{, which features preconditioned Krylov subspace
  methods,} and perform official weak scalability tests in the TOP500
supercomputer list.  \checkPM{Finally, it is also worth noting that we
  are planning to port HLAM to accelerators such as GPUs and FPGAs by
  making use of third-party libraries or ad hoc kernels.}

\section*{Acknowledgements}
\label{sec:acknowledgements}
This work has received funding from the European High Performance
Computing Joint Undertaking (EuroHPC JU) initiative [grant number
  956416] via the exaFOAM research project. The JU receives support
from the European Union's Horizon 2020 research and innovation
programme and France, Germany, Spain, Italy, Croatia, Greece, and
Portugal. In Spain, it has received complementary funding from
MCIN/AEI/10.13039/501100011033 [grant number PCI2021-121961]. This
work has also benefited financially from the Ram\'on y Cajal programme
[grant number RYC2019-027592-I] funded by MCIN/AEI and
ESF/10.13039/501100004895 as well as the Severo Ochoa Centre of
Excellence accreditation [grant number CEX2021-001148-S] funded by
MCIN/AEI. The Programming Models research group at BSC-UPC received
financial support from Departament de Recerca i Universitats de la
Generalitat de Catalunya [grant number 2021 SGR 01007].

\section*{Glossary}
\label{sec:glossary}
\begin{description}
  \item[BiCGStab] biconjugate gradient \checkPM{stabilized}
  \item[BiCGStab-B1] biconjugate gradient \checkPM{stabilized}, one blocking
  \item[CFD] computational fluid dynamics
  \item[CG] conjugate gradient
  \item[CG-NB] conjugate gradient, nonblocking
  \item[GS] Gauss--Seidel
  \item[HDOT] hierarchical domain overdecomposition with tasks
  \item[HLAM] hybrid linear algebra methods
  \item[HPCCG] high performance computing conjugate gradients
  \item[KSM] Krylov subspace method
  \item[MPI] message passing interface
  \item[OMP] OpenMP
  \item[OSS] OmpSs-2
  \item[SpMV] sparse matrix--vector multiplication
  \item[TAMPI] task-aware message passing interface
\end{description}

\appendix

\section{Code snippets}
\label{sec:code-snippets}

\begin{lstfloat}
  \lstinputlisting[caption={First tasks inside the main loop of the
      CG-NB algorithm.}, label={code:CG-NB}]{\codeSrc{CG-NB.cpp}}
\end{lstfloat}
\begin{lstfloat}
  \lstinputlisting[caption={Point-to-point communication tasks embedded in HLAM.}, label={code:MPI}]{\codeSrc{MPI.cpp}}
\end{lstfloat}
\begin{lstfloat}
  \lstinputlisting[caption={Fork-join parallelism implemented within the SpMV kernel.}, label={code:SpMV}]{\codeSrc{SpMV.cpp}}
\end{lstfloat}
\begin{lstfloat}
  \lstinputlisting[caption={Computation tasks inside the main loop of
      the relaxed, symmetric Gauss--Seidel algorithm.},
    label={code:relaxed-Gauss-Seidel}]{\codeSrc{relaxed-Gauss-Seidel.cpp}}
\end{lstfloat}
%

%%       \citet{<label>} ==> Jones et al. [21]
%%       \citep{<label>} ==> [21]
%%

\bibliographystyle{elsarticle-num-names}
\bibliography{HLAMpaper.bib}


%% \begin{thebibliography}{00}


%% \bibitem[ ()]{}

%% \end{thebibliography}
\end{document}